\documentclass[%
 reprint,
superscriptaddress,
nofootinbib,
 amsmath,amssymb,
 aps,
prd,
longbibliography
]{revtex4-1}

\usepackage[utf8]{inputenc}
\usepackage{graphicx}
\usepackage{dcolumn}
\usepackage{bm}
\usepackage{hyperref}
\usepackage{color}
\usepackage[mathlines]{lineno}
\usepackage{tabularx}
\usepackage[normalem]{ulem}
\usepackage{gensymb}
\usepackage[export]{adjustbox}
\usepackage{url}
\usepackage{mathtools} 
\usepackage{float}
\usepackage{ amssymb }
\hypersetup{
    pdfnewwindow=true,    
    colorlinks=true,      
    linkcolor=blue,       
    citecolor=blue,       
    filecolor=blue,       
    urlcolor=blue         
}

\newcolumntype{Y}{>{\centering\arraybackslash}X}

\DeclareRobustCommand{\VAN}[3]{#2}
\let\VANthebibliography\thebibliography
\def\thebibliography{\DeclareRobustCommand{\VAN}[3]{##3}\VANthebibliography}

\setcitestyle{authoryear}

\usepackage{natbib}

\begin{document}

\author{Kelvin H. M. Chan}
\email{kelvinhmchan@link.cuhk.edu.hk}
\thanks{\scriptsize \!\! \href{https://orcid.org/0000-0002-2019-2025}{orcid.org/0000-0002-2019-2025}}
\affiliation{Department of Physics, The Chinese University of Hong Kong, Shatin, New Territories, Hong Kong}

\author{Otto A. Hannuksela}
\email{otto.akseli.hannuksela@gmail.com}
\thanks{\scriptsize \!\! \href{https://orcid.org/0000-0002-3887-7137}{orcid.org/0000-0002-3887-7137}}
\affiliation{Department of Physics, The Chinese University of Hong Kong, Shatin, New Territories, Hong Kong}

\date{ \today}

\title{Extracting ultralight boson properties from boson clouds around post-merger remnants}

\begin{abstract}
\noindent 
Ultralight bosons are a class of hypothetical particles that could potentially solve critical problems in fields ranging from cosmology to astrophysics and fundamental physics. If ultralight bosons exist, they form clouds around spinning black holes with sizes comparable to their particle Compton wavelength through superradiance, a well-understood classical wave amplification process that has been studied for decades. After these clouds form, they dissipate and emit continuous gravitational waves through the annihilation of ultralight bosons into gravitons. These gravitons could be detected with ground-based gravitational-wave detectors using continuous-wave searches. However, it is conceivable for other continuous-wave sources to mimic the emission from the clouds, which could lead to false detections. Here we investigate how one can use continuous waves from clouds formed around known merger remnants to alleviate this problem. In particular, we simulate a catalogue of merger remnants that form clouds around them and demonstrate with select "golden" merger remnants how one can perform a Bayesian cross-verification of the ultralight boson hypothesis that has the potential to rule out alternative explanations. Our proof-of-concept study suggest that, in the future, there is a possibility that a merger remnant exists close enough for us to perform the analysis and test the boson hypothesis if the bosons exist in the relevant mass range. Future research will focus on building more sophisticated continuous-wave tools to perform this analysis in practice. 
\end{abstract}

\maketitle

\section{\label{sec:intro} Introduction}
Ultralight bosons (ULBs) are a broad class of hypothetical integer spin particles with mass $\ll$ 1eV \cite{PhysRevD.16.1791, PhysRevLett.40.223, PhysRevLett.40.279}, including axion-like particles \cite{axion1983}, dilatons \cite{dilatons2015}, and Majorons~\cite{Chikashige1981265}. They have been promising dark matter (DM) candidates \cite{dm2005,Ackerman:2008kmp,Fairbairn:2014zta,Marsh:2014qoa,Marsh:2015xka,dm2018,Braaten:2019knj} and solutions to various problems in particle physics, astrophysics, and cosmology\cite{dm2005,Herizberg2008, Arvanitaki:2009fg,Asztalos:2009yp, Arvanitaki2011, Kim:2015yna, dilatons2015, HUI2017, Arvanitaki2020, Mehta:2020kwu, wavedm2021}. 

If ULBs exist in nature, they will couple to rotating black holes with Schwarzschild radii similar to the boson's compton wavelength through a process called superradiant instability, forming clouds of boson particles~\citep[e.g.][]{Arvanitaki:2009fg,Brito:2015oca}. 
In particular, when a superradiant instability is spontaneously triggered, the boson fields around a rotating black hole (BH) will extract energy and angular momentum of the host BH to create bosons around the BH~\cite{damour1976quantum, zouros1979instabilities, Brito:2015oca, Baumann_2019, Isi2019, Baryakhtar:2017ngi, Tensor_2020}. 
The process continues to spin down the host BH and extract its mass and angular momentum until the boson Compton frequency matches approximately the rotational frequency of the black hole~\citep{Brito:2015oca}. 
The energy and angular momentum is used to form a boson cloud around the BH. 
Indeed, the final product is a black hole/cloud system. 

It is particularly interesting that the formation of such a bosonic cloud could lead to potentially observable signatures. 
In particular, when the cloud reaches a macroscopic size, it can emit continuous quasimonochromatic GWs through annihilation of bosons into gravitons \cite{Arvanitaki2011, Arvanitaki:2014wva, Brito:2017zvb, Isi2019, Tsukada:2018mbp,Brito:2017wnc,Fan:2017cfw}. 
Furthermore, because the superradiance effect spins down rotating black holes, another observational signature is a dearth of rotating black holes above the expected critical spin set by the boson mass (so-called "Regge trajectory")~\citep{Brito:2015oca}. 
Finally, it is also possible for the presence of the cloud to disturb binary orbits~\citep{Ferreira:2017pth,Baumann2019,Hannuksela:2018izj,Baumann:2019ztm,Zhang:2018kib,Bar:2019pnz,Berti:2019wnn,Amorim:2019hwp,Cardoso:2020hca} and, under specific scenarios, couple to magnetic fields and emit photons~\citep{Boskovic:2018lkj,Day:2019bbh} or explode in a "bosenova" due to self-interactions~\citep{Brito:2015oca, East:2022ppo}.\footnote{Note that the mechanism behind the scalar version of the bosenova in the non-relativistic regime has been discussed in the recent literature~\citep{Baryakhtar:2020gao}.} 
Indeed, a wide range of observational signatures are possible.  

Scientists can now utilize a wide range of astrophysical messengers, including the wide spectrum of electromagnetic waves, cosmic rays,  and neutrinos in the hunt for beyond-standard-model particles. 
In the recent years, gravitational-wave observations have also become commonplace~\citep{GWTC3}. 
With the advent of these new gravitational observations, scientists can now perform searches for DM candidates like primoridial black holes and ULBs using GWs in addition to the existing messengers \cite{gwdm2020, o3_cloud, o3a_pbh, LIGOScientificCollaborationVirgoCollaboration:2021eyz}. 

In particular, several efforts search for these ultralight bosons using table-top experiments or astronomical observations~\cite{Asztalos:2009yp,Wagner:2010mi,Rybka:2010ah,Arik:2011rx,Pugnat:2013dha,Arvanitaki:2014faa,Corasaniti:2016epp,Choi:2017hjy,Akerib:2017uem,Brubaker:2016ktl,Kim:2017yen,Garcon:2017ixh,Arvanitaki:2014wva,Arvanitaki:2016qwi,Baryakhtar:2017ngi,Brito:2017wnc,Cardoso:2018tly,antonio2018cw,Ghosh:2018gaw,Tsukada:2018mbp,Stott:2017hvl,Hannuksela:2018izj,Ouellet:2018beu,Davoudiasl:2019nlo,Fernandez:2019qbj,Palomba:2019vxe,Ng:2020jqd,Abel:2017rtm,Grote:2019uvn,Dev:2016hxv,Zhu:2020tht,Ng:2020ruv}. 
Spin measurements of BHs in X-ray binaries~\cite[see][]{Remillard:2006fc,Middleton:2015osa} could be used to search for bosons in the mass ranges of $5~\mathrm{M}_{\odot}< M\lesssim 20~\mathrm{M}_{\odot}$~\cite{Remillard:2006fc,Corral-Santana:2015fud}. 
Gravitational-wave observations of black hole spins from binary black hole coalescences provide another avenue, as they encode the properties of their sources, including the masses and spins of the two component BHs, which could allow us to observe the dearth of high-spin black holes predicted by the existence of the ULBs~\citep{Brito:2017zvb, Ng:2019jsx,Ng:2020ruv}. 
Since ground-based GW detectors can detect heavier BHs ($M$ up to $\sim100~\rm{M}_{\odot})$~\cite{GWTC1,GWTC2,GWTC3} than those found in X-ray binaries, the spin measurements inferred from GWs probe a lighter range of boson mass. 
Another major, related area is to study signatures in the observations of gravitational waveforms during binary coalescence \cite{Yang:2017lpm, Baumann2019, Annulli_2020, Kavanagh_2020, Chung:2021roh}. 
Finally, existing searches also include directed searches and all-sky search for continuous gravitaional waves (CWs). 
These CW searches employ semi-coherent methods such as the Hidden Markov Model (HMM), in which the full-length data are broken into smaller chunks of data to analyze coherently \cite{antonio2018cw, Isi2019, Palomba:2019vxe, o3_cloud} (see \cite{Riles:2017evm} for a review on CW searches). 

However, there are certain known limitations to the existing searches. 
For example, searching for holes in the Regge plane requires the timescale between the formation of BHs and the merger to be larger than the superradiant timescale for the ULB cloud to form and spin down the BH~\cite{Brito:2017zvb, Ng:2019jsx,Ng:2020ruv}, and the formation or presence of the clouds could in theory be tidally disrupted~\citep{Baumann:2019ztm,Berti:2019wnn}. 
In addition, searching for the CW signals from ULB clouds may be difficult as the other CW sources could in principle mimic the GWs from a ULB cloud. 

One way to reduce the probability of false detections in CW searches is with directed searches targeting known merger remnants~\citep{Arvanitaki:2014wva,Arvanitaki:2016qwi,Baryakhtar:2017ngi,Isi2019,Ghosh:2018gaw,Ng:2020jqd}.
Intuitively, we know the sky position of any observed merger remnant, and therefore a targeted search would both allow us to reduce the background noise but also link the CW signal to the merger remnant, reducing the probability that the signal originated from another source~\citep{Isi2019,Ghosh:2018gaw,Ng:2020jqd}. 
In this work, we demonstrate how the remnant BH's properties can, in addition, be used in the analysis to further reduce the probability of false detections. 

In particular, we know the mass and spin of any observed merger remnant. 
Together with the CW observation, the remnant properties would allow us to theoretically link the CW emission to the BH properties. 
If the prediction matches the actual observation, we can conclude that the signal originates from an ultralight boson at very high confidence. 
Indeed, we demonstrate that one could robustly rule out other CW sources that are unable to produce a consistent signal using the merger remnant information.

In this work, we focus on scalar (spin-0) and vector (spin-1) bosons, where the physics is well-studied \cite{Arvanitaki2011,vector_pani2012,vector_rosa2012,Brito:2017zvb,Baryakhtar:2017ngi,vector_east2017,Isi2019, Baumann2019, Baumann_2019, Brito:2015oca, Brito:2017zvb}. Meanwhile, theoretical studies on tensor (spin-2) bosons \cite{tensor2013, Tensor_2020} have yet to include a comprehensive calculation on gravitational-wave emissions. 

The paper is structured as follows: 
in Sec.\ref{sec:boson_cloud}, we review the physics of the ULB clouds formed around rotating BHs and the GW emission. 
We then discuss how focusing on clouds around merger remnants may aid the searches for ULBs. In Sec.\ref{sec:horizon}, we use the standard signal-to-noise (SNR) calculations in gravitational-wave physics to evaluate the detection horizon of clouds formed around merger remnants by different GW observatories. 
We then compare the horizon with the existing forecast on merger events. 
In Sec. \ref{sec:analysis}, we perform a mock analysis with a 'golden' remnant from the merger forecast catalogue, to show that by obtaining a coherent boson cloud measurement, we can cross-verify the existence of ULBs. 
Finally, we conclude in Sec.\ref{sec:conclusions}.

\begin{figure*}[t]
\centering
\includegraphics[width=\linewidth]{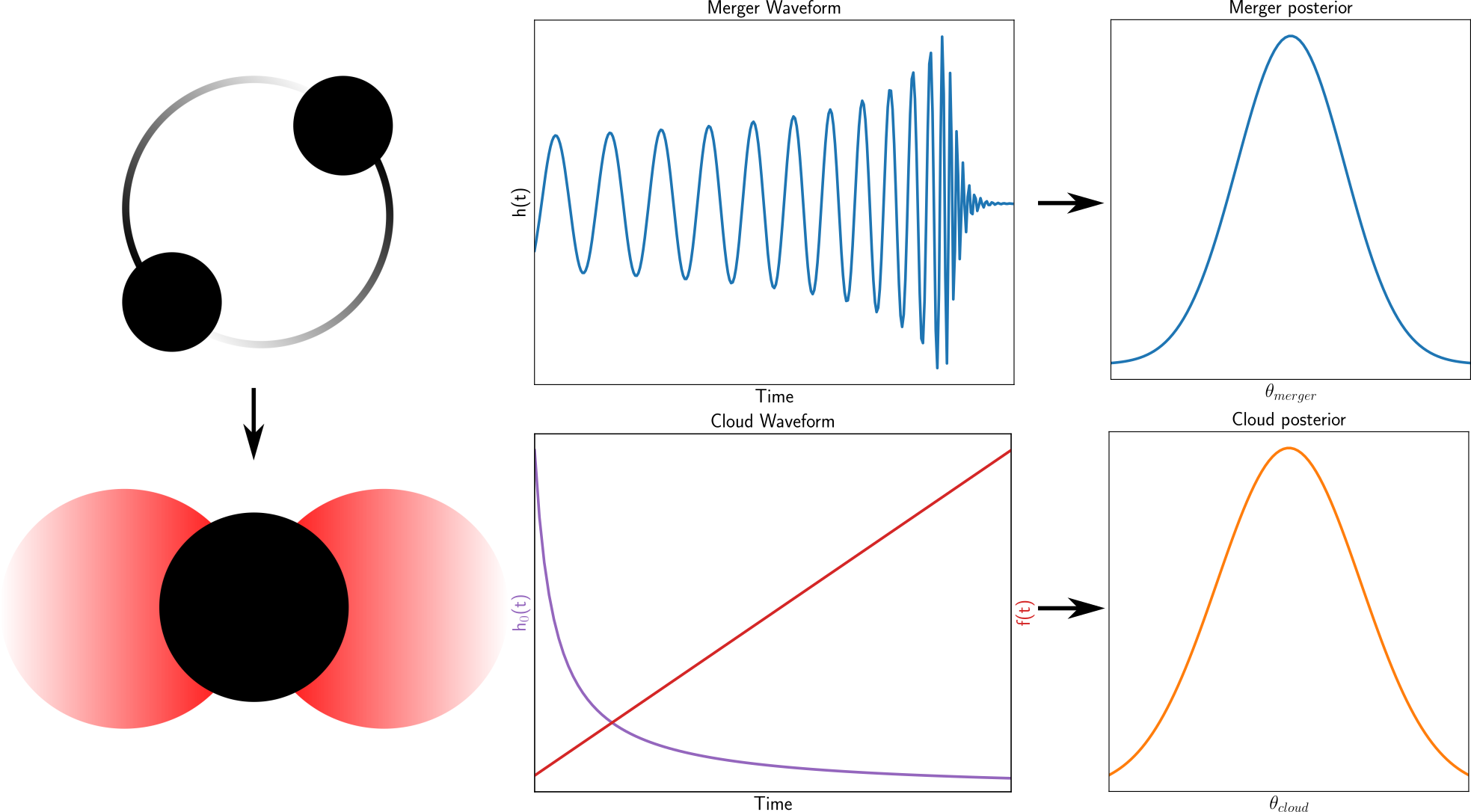}
\caption{
An illustration of the search for ULBs using post-merger remnants. \textit{Top:} When a system of binary black holes merges, it radiates gravitational waves, from which we can infer the parameters $\theta_{merger}$ of the binary black hole system. \textit{Bottom:} Due to superradiance, ultralight bosons form clouds around the rotating black hole merger remnant by extracting energy and angular momentum from it. As the mass of the boson cloud grows by spinning down the BH, the bosons simultaneously and with increasing efficiency annihilate to emit gravitational waves. We can infer the parameters $\theta_{cloud}$ of the black hole-boson cloud system through studying these gravitational waves. Combining both measurements, one could obtain robust evidence of the existence of ultralight bosons.}
\label{fig:illustration}
\end{figure*}

\begin{figure*}[t]
\centering
\includegraphics[width=0.49\textwidth]{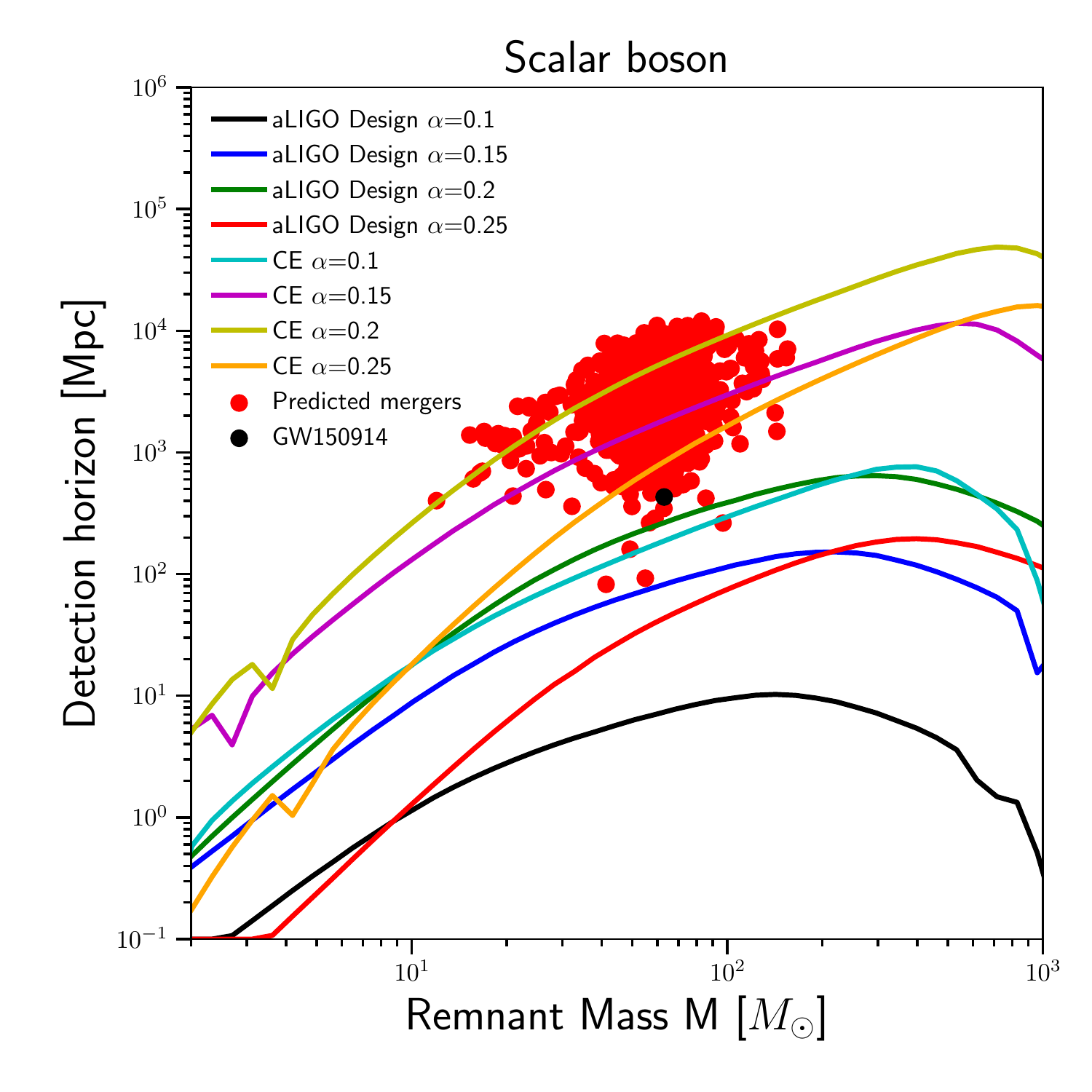}
\includegraphics[width=0.49\textwidth]{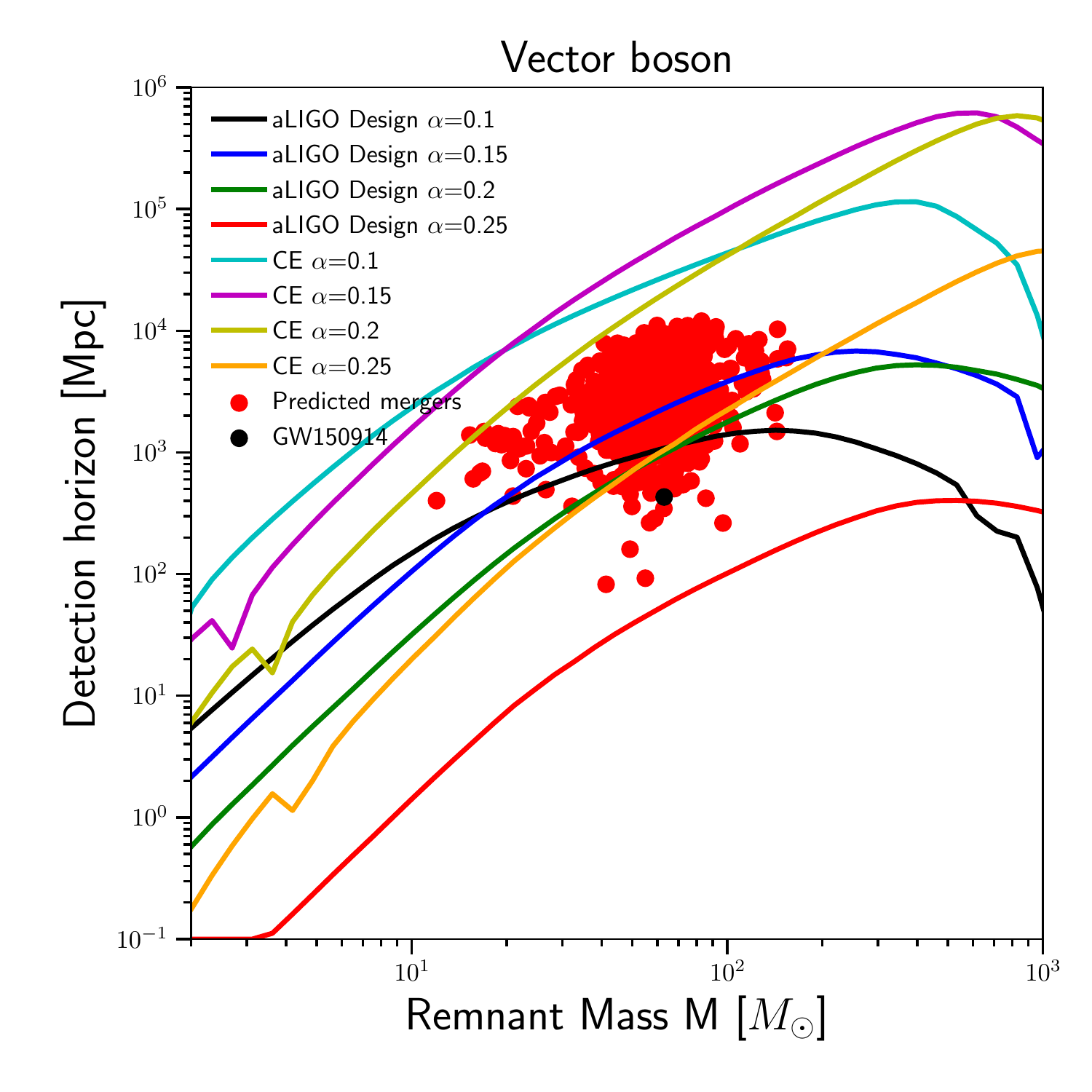}
\caption{Detection horizon of ULB cloud formed around merger remnants vs host BH mass by different detector, along with the scatter of predicted and the observed binary BH merger event GW150914. The parameter space under the curve would be detectable by the respective detector. We take the initial spin being 0.7, a value similar to most known merger remnants. We choose several $\alpha$ values that would satisfy the superradiance condition specified in Eq. \ref{superradiance_condition}. We assume 3 years of observation time for scalar or 1 day of observation time for vector. For both scalar and vector, four 'golden' merger remnants are predicted to host ULB cloud within the horizon of the LIGO detectors at design sensitivity.}
\label{fig:horizon}
\end{figure*}

\begin{figure}
\centering
\includegraphics[width=\columnwidth]{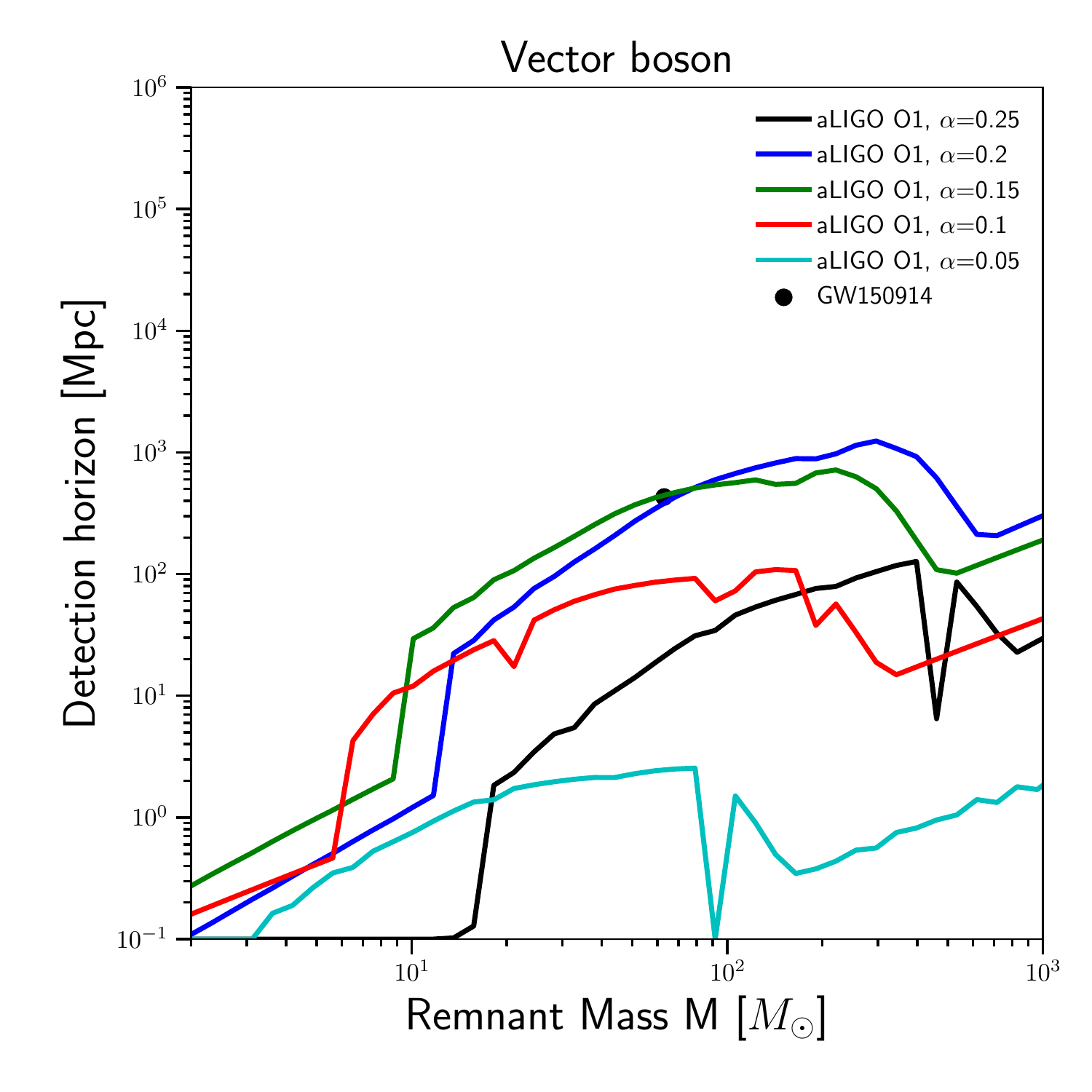}
\caption{Same horizon plot as in Fig. \ref{fig:horizon}, but for the first run O1 \cite{LIGO_O1} sensitivity. O1 sensitivity could be marginally detecting the vector boson cloud hosted by the remnant from the first gravitational-wave observation GW150914 \cite{ligo2016}, if the boson mass falls into the right range ($\alpha\sim$0.15-0.2, which corresponds to a boson mass $\mu$ of $3.2 \times 10^{-13} - 4.3 \times 10^{-13}$ $\rm eV$). }
\label{fig:horizon_o1}
\end{figure}

\begin{figure*}
\centering
\includegraphics[width=0.49\textwidth]{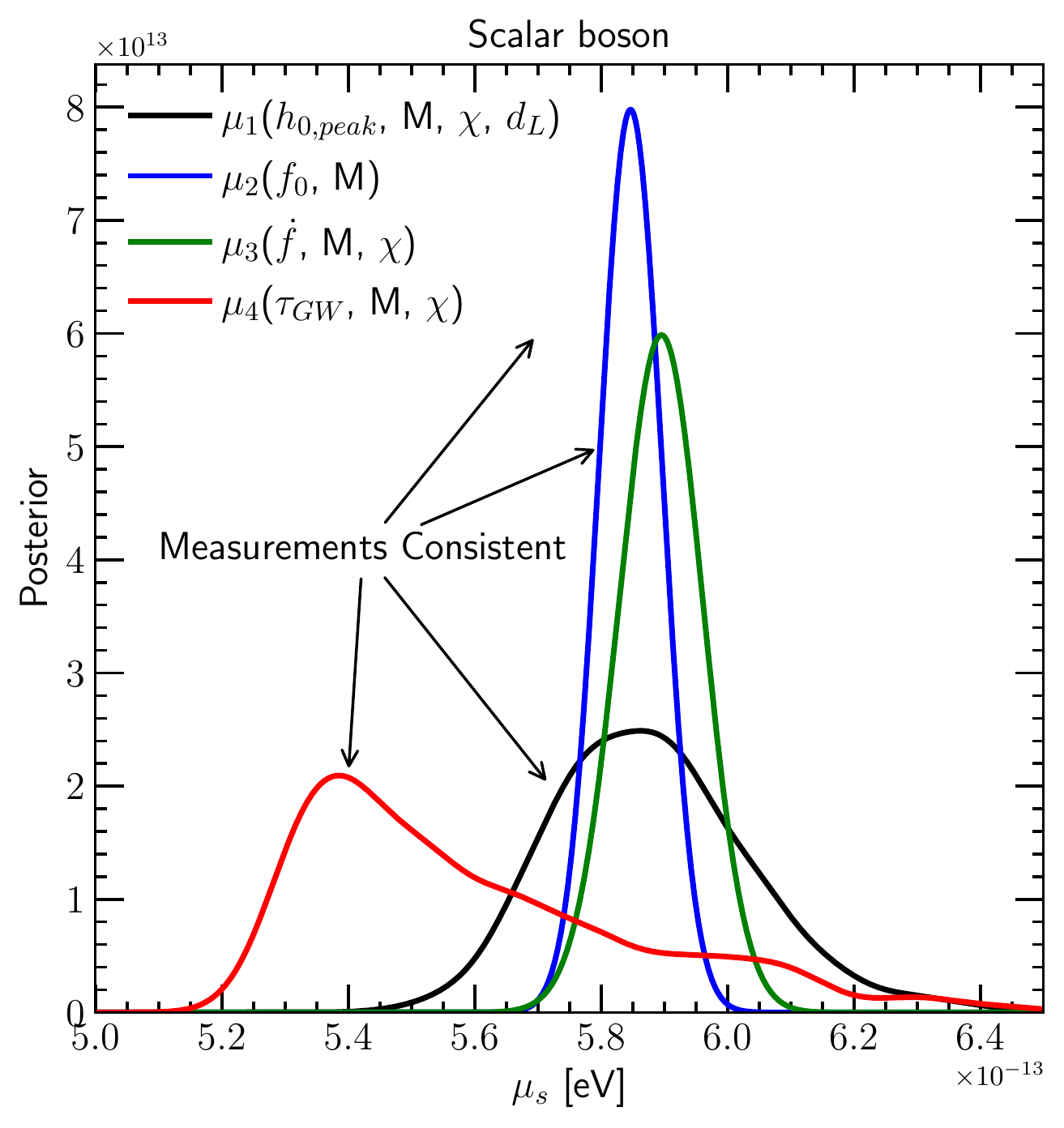}
\includegraphics[width=0.49\textwidth]{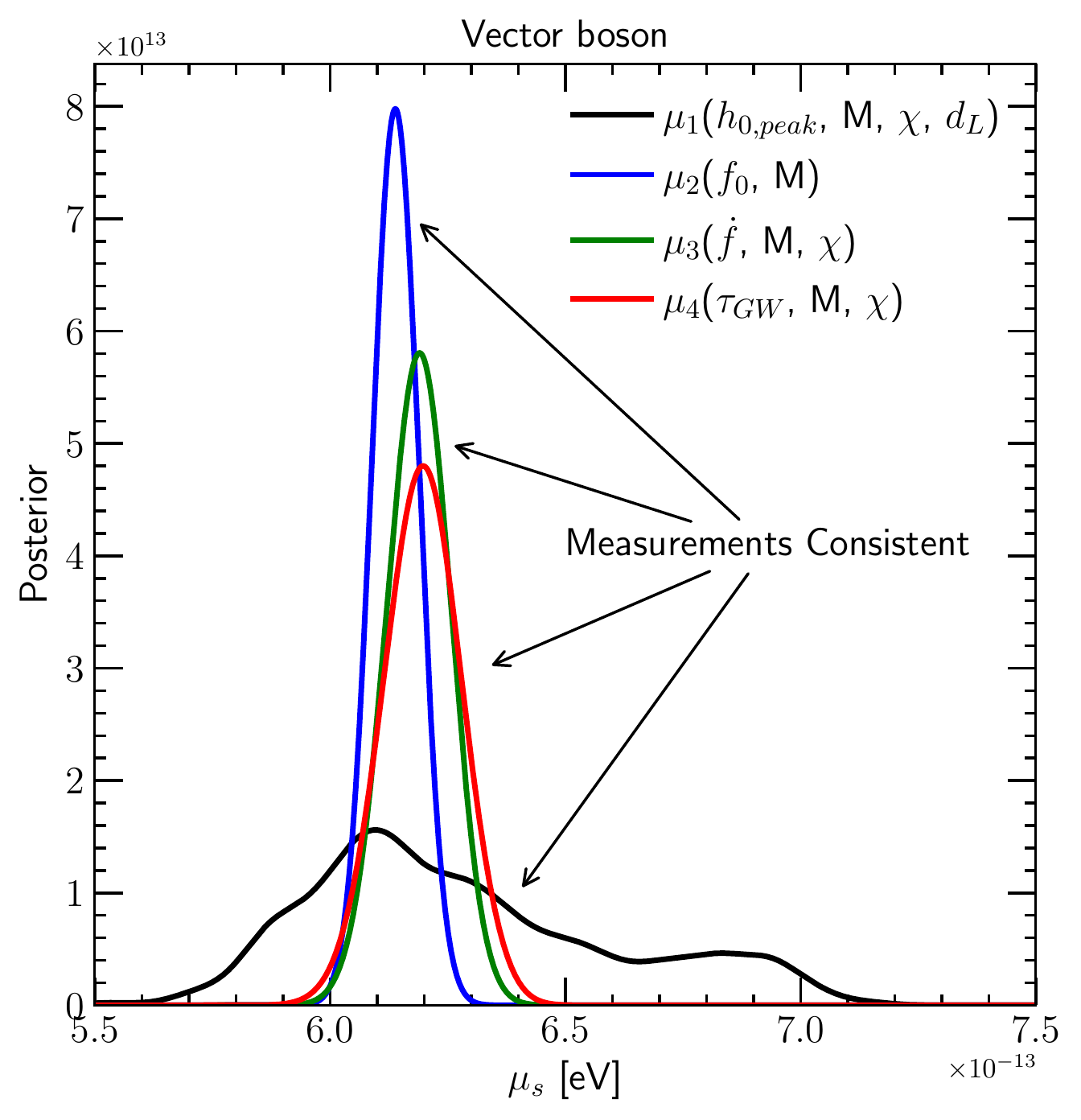}
\caption{An example of the posterior. If ultralight boson exists in nature, the four different measurements of the gravitational waves emitted by the cloud would provide an agreed value of boson mass. Here, we use the binary black hole merger remnant system with properties listed in Table \ref{table:remnants_properties} and $\alpha$=0.2 to simulate gravitation waves signal release by the merger and the ultralight boson cloud formed around the remnant. The SNR is 11.42 for the scalar cloud (left) and 46.06 for the vector cloud (right). We then perform parameter estimation using \texttt{Bilby}~\cite{bilby2019} to get the posterior from measurements of the merger and the cloud independently, which are then combined to obtain the posterior of boson mass. It shows four agreed boson mass posterior, signifying the existence of ultralight boson.}
\label{fig:para_est}
\end{figure*}

\begin{figure}
\centering
\includegraphics[width=\columnwidth]{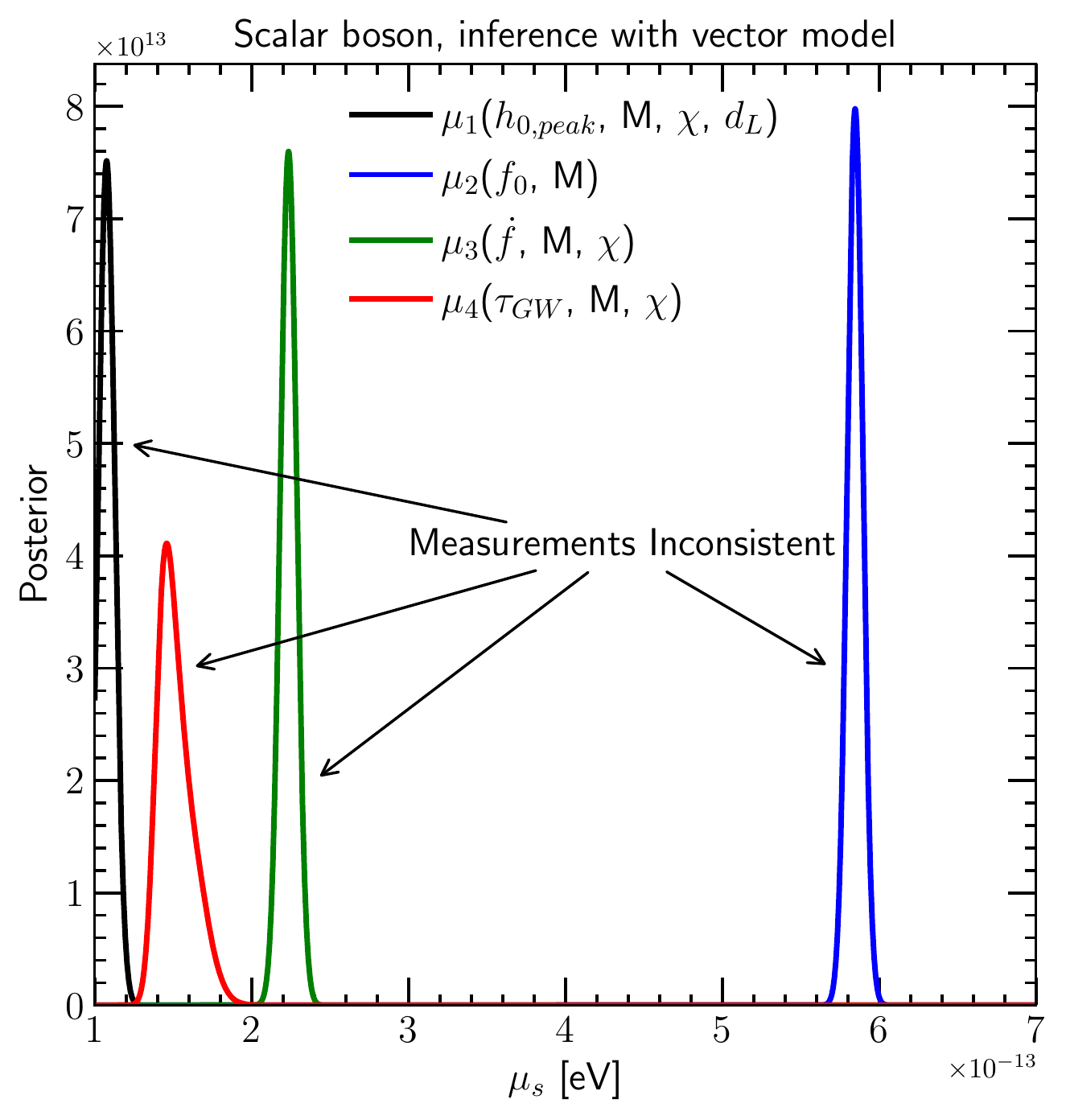}
\caption{Disentangling the signal between scalar and vector boson. To do so, we inject GW signal from scalar boson and inference the boson mass with vector model. The four measurements of the boson mass will not be consistent as in Fig. \ref{fig:para_est}. Thus we can distinguish whether the signal is from a scalar boson cloud or a vector boson cloud.}
\label{fig:disentangle}
\end{figure}

\begin{figure}
\centering
\includegraphics[width=\columnwidth]{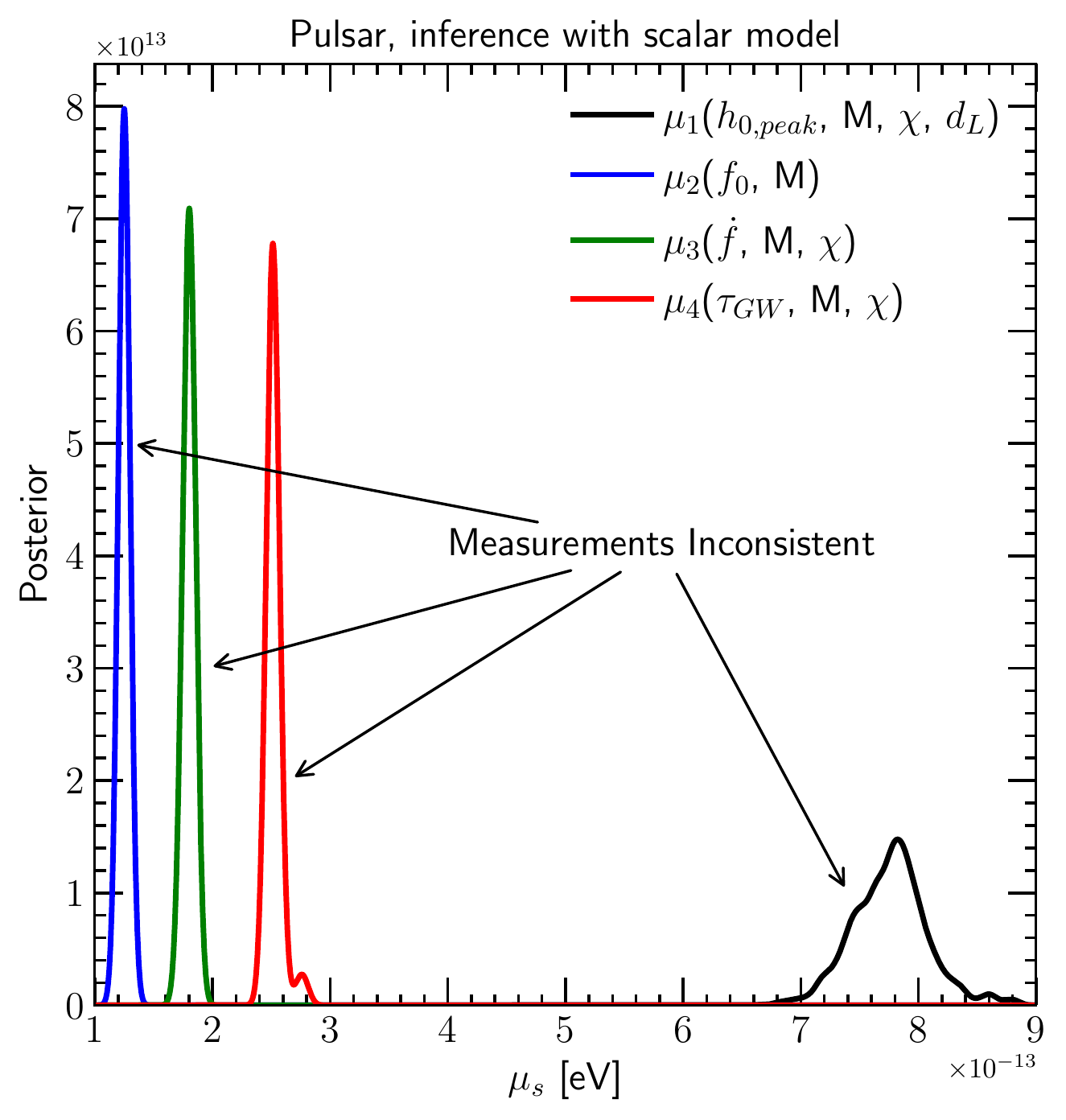}
\caption{The analysis can be used to distinguish CW signals from ULB cloud and other sources. We inject GW signal from pulsar and inference the boson mass. Four measurements of the boson mass are not consistent, hence we can distinguish whether the signal is from ULB cloud or other sources.}
\label{fig:pulsar}
\end{figure}

\textbf{\section{\label{sec:boson_cloud} The boson cloud}}

In this section, we review the physics of boson fields around rotating BHs in a manner that is best suited for the later discussion. We first review how macroscopic boson clouds spontaneously form, and then the gravitational waves emission by the cloud.\par\medskip

\subsection{\label{subsec: CloudFormation}Cloud formation}

For a Kerr BH \cite{KerrBlackHole} of mass M and dimensionless spin $\chi$, we can define the characteristic length \cite{Schwarzchild1,Schwarzchild2}
\begin{equation}
 r_g=GM/c^2   
\end{equation}

\noindent which is half the Schwarzchild radius $r_s$. The radius of the BH's outer horizon is \cite{Boyer-Linquist}
\begin{equation}
r_+=r_g\bar{r}_+=r_g(1+\sqrt{1-\chi^2}).
\end{equation}

At the outer horizon, the frame-dragging angular velocity is \cite{Teukolsky_Kerr}
\begin{equation}
    \Omega_{BH} = \frac{1}{2}\frac{c}{r_g}\frac{\chi}{\bar{r}_+}.
\end{equation}

If beyond the Standard Model particles, there exist ULBs with mass \cite{PhysRevD.16.1791,PhysRevLett.40.223,PhysRevLett.40.279}
\begin{equation}
    m_b=\mu/c^2,
\end{equation}
\noindent where $\mu$ is the rest energy, the angular frequency corresponding to the Compton wavelength $\lambdabar$ is \cite{ComptonWavelength}
\begin{equation}
\omega_{\mu}=c/\lambdabar=\mu/\hbar.
\end{equation}

When a BH is born, the quantum fluctuations will lead to the pair production of particles in the vicinity of the BH. 
If the wavelength of the particles $\lambdabar$ is comparable to the Schwarzchild radius of the BH $r_s$, this instability quickly extracts energy and angular momentum from the BH to increase the number of particles, a process called "superradiant instability." Such a process occurs when the "superradiance condition" is satisfied \cite{Superradiance_condition}:
\begin{equation}
\label{superradiance_condition}
    \omega_{\mu}/m<\Omega_{BH},
\end{equation}
\noindent where m is the magnetic quantum number, which is the projection of the particle's total angular momentum to the BH spin direction. 
As the field has non-zero mass, the bosons are gravitationally bounded to the BH. 
The superradiance process continues to occur until Eq. \ref{superradiance_condition} is no longer satisfied. 
The cloud can extract at most $\sim 10\%$ of the BH's mass~\citep{East:2017ovw}.

Interestingly, the structure of the cloud as well as the cloud solution is very similar to the electron cloud solution around a hydrogen atom~\citep[e.g.][]{Arvanitaki:2009fg,Brito:2015oca}.
For this reason, it is convenient to define a so-called gravitational fine-structure constant $\alpha$, which plays the same role as the fine-structure constant in the hydrogen atom, and takes the value of the ratio of the two length scales \cite{Detweiler:1980uk, Dolan_2007}:
\begin{equation}
    \alpha=\frac{r_g}{\lambdabar}=\frac{\frac{GM}{c^2}}{\frac{\hbar c}{\mu}}=\frac{G}{c^3\hbar}M\mu.
\end{equation}

\subsection{\label{subsec: GW} Gravitational-wave emission by the boson cloud}

Once the boson cloud is of a macroscopic size, it can emit gravitational radiation through three mechanisms \cite{Isi2019}: 
(i) annihilation of bosons into gravitons; 
(ii) bosenova: the supernova-like collapse of the cloud due to boson self-interactions; and 
(iii) boson transitions between energy levels, analogous to electrons in the hydrogen atom. 
Bosenova signals (ii) last order of milliseconds~\citep{Brito:2015oca,Baryakhtar:2020gao, East:2022ppo}, making bosenovae a better target for burst-like searches rather than continuous searches. 
The boson transitions between energy levels (iii) occurs in very old BHs, but the remnants that we target are new-born black holes, which makes observations of these transitions unlikely in young merger remnants. 
Thus, here we focus on the first of the three mechanisms, annihilation of bosons into gravitons.

Hence, as in \cite{Isi2019}, in this work, we restrict ourselves to signals from annihilation only.   
We focus on the dominant level cloud, where the instability timescale is the shortest. 
For dominant scalar cloud $(n, l, j, m)=(2, 1, 1, 1)$, and for dominant vector cloud $(n, l, j, m)=(1, 0, 1, 1)$ where n, j, m are the usual quantum numbers of the hydrogen atom. 
Including the higher order modes may become important in a realistic search, but we expect that they would primarily introduce second-order corrections in our mock data analysis.
The corrections could boost the signal strength and allow us to probe ultralight boson clouds slightly farther away and also enable one to perform a more rigorous parameter estimation for the ultralight bosons that could also check test for jumps in the GW frequency caused by a jump in the cloud mode. 

Here we follow the approximations by \cite{Arvanitaki2011,Baryakhtar:2017ngi,Isi2019,Zhu2020}, except for the gravitational-wave amplitude for the scalar boson, for which we adopt the non-relativistic approximation. The boson cloud would emit quasi-monochromatic GWs. The initial frequency of the CWs emitted by the boson cloud is 
\begin{equation}
    \label{eq:f}
    f_0 \approx 645 \textrm{Hz} \left( \frac{M}{10M_{\odot}} \right) \left( \frac{\alpha}{0.1} \right) .
\end{equation}
After formation, the frequency would increase over time since cloud mass decreases. 
The frequency drift is approximated by
\begin{equation}
    \label{eq:df_s}
    \dot{f}^{(s)} \approx 3 \times 10^{-14} \textrm{Hz/s} \left( \frac{M}{10M_{\odot}} \right)^{2} \left( \frac{\alpha}{0.1} \right)^{19} \chi^{2}  \,,
\end{equation}
for the dominant scalar mode and 
\begin{equation}
    \label{eq:df_v}
    \dot{f}^{(v)} \approx 1 \times 10^{-6} \textrm{Hz/s} \left( \frac{M}{10M_{\odot}} \right)^{2} \left( \frac{\alpha}{0.1} \right)^{15} \chi^{2}  
\end{equation}
for the dominant vector mode. 
The frequency drift is much faster for vector bosons than for scalar bosons, because vector clouds emit gravitational waves at a faster rate.

When the BH has been fully spun down, the gravitational-wave strain amplitude emitted by the dominant mode boson cloud surrounding the BH is approximated by \cite{Isi2019}
\begin{equation}
    h^{(s)}_{0, peak} \approx 8 \times 10^{-28} \left( \frac{M}{10M_{\odot}} \right) \left( \frac{\alpha}{0.1} \right)^{7} \left( \frac{Mpc}{d_L} \right) \left( \frac{\chi - \chi_{f}}{0.1} \right) \,,
\end{equation}
for the dominant scalar mode and 
\begin{equation}
    \label{eq:h_v}
    h^{(v)}_{0, peak} \approx 4 \times 10^{-24} \left( \frac{M}{10M_{\odot}} \right) \left( \frac{\alpha}{0.1} \right)^{5} \left( \frac{Mpc}{d_L} \right) \left( \frac{\chi - \chi_{f}}{0.1} \right)\,,  
\end{equation}
for the dominant vector mode. 
The strain of the annihilation is largest when the cloud first reaches maximum occupation number. As the bosons annihilate and deplete the cloud, the strain amplitude $h_0(t)$ decreases over time
\begin{equation}
    \label{eq:dh}
    h_{0}(t) \approx \frac{h_0}{1+\frac{t}{\tau_{GW}}}
\end{equation}
where $\tau_{GW}$ is the gravitational-wave timescale, which is the time it takes to radiate away half of the cloud's rest energy:
\begin{equation}
    \label{eq:tau_gwsv}
\begin{split}
    \tau^{(s)}_{GW} &\approx 6.5 \times 10^{4} \textrm{yr} \left( \frac{M}{10 M_{\odot}} \right) \left( \frac{0.1}{\alpha} \right)^{15} \frac{1}{\chi}\,. \\
    \tau^{(v)}_{GW} &\approx 1 \textrm{day} \left( \frac{M}{10 M_{\odot}} \right) \left( \frac{0.1}{\alpha} \right)^{11} \frac{1}{\chi}\,.
\end{split}
\end{equation}
Since the frequency evolution is slow, we approximate the gravitational waves emitted by the boson cloud to take the waveform $h(t)$ expanding the time-evolution of the frequency in its first derivative only, such that 
\begin{equation}
    \label{eq:h_t}
    h(t) = h_{0, peak}(t) e^{i \omega t} = \frac{h_{0, peak}}{1+\frac{t}{\tau_{GW}}} e^{i2\pi (f_0+ \dot{f}t)t}.
\end{equation}
where the angular frequency slowly evolves with time: $\omega = 2\pi (f_0+ \dot{f}t)$.
Including  higher-order derivatives or formulating a more complete waveform would allow for more accurate results. 
In particular, the non-relativistic expression for the GW amplitude in the scalar and vector case is accurate within approximately an order of magnitude.  Specifically,~\citet{vector_east2017, Brito:2017zvb, East:2018glu} find the amplitude using full GR time-domain simulation may differ for the loudest signals by about an order of magnitude \cite{Siemonsen:2019ebd}. 
A similar correction may result in the gravitational-wave frequency drift. 
However, the principle test behind our analysis would be unchanged by such corrections. 
\subsection{\label{subsec: remnant}Post-merger Remnants}

In the following sections, we will limit ourselves to clouds formed around binary BH merger remnants. 
It is advantageous to perform searches targeting post-merger remnants for the following reasons:
\begin{enumerate}
    \item By analyzing the GWs from the mergers, we know the location of the remnants. In addition to the three-detector network of LIGO Hanford, Livingston, and Virgo, KAGRA has become operational \cite{KAGRA:2020tym, KAGRA2013} and LIGO-India \cite{LIGOindia_proposal, INDIGO2013} is expected to become online in the coming years. Thus, in the future, we expect to be able to better localize merger events. Hence, after we detect signals from mergers,  we can perform follow-up directed searches for ultralight boson clouds in the direction of the merger remnant~\citep{Isi2019}.
    \item By analysing the mergers waveform, we have information about BH mass $M$, BH spin $\chi$ and luminosity distance $d_L$. From these, and combining the CW waveform, we can infer the boson mass $\mu$.
    \item There are four measurement depending on $\mu$: $f$, $\dot{f}$, $h$, $\dot{h}$. As a result, we can infer four $\mu$ from these four measurements, which can provide robust evidence for the existence of ULBs: if four measurements $\mu$ agree, it is very likely that ULBs exist, as there is no other reasons for all the $\mu$ to agree; if four measurements of $\mu$ differs, we can rule out the existence of ULBs at a particular mass range.
    \item Most importantly, we can rule out other possible CW sources, such as neutron stars, as it is unlikely that the signal is given by other CW sources if the measurements of boson properties agree.
\end{enumerate}

\section{\label{sec:horizon} Detection Horizon}

\subsection{Horizon calculation}
Using the approximation in Sec. \ref{subsec: GW}, we can find the detection horizon, where the signal is barely detectable, of ULB clouds formed around post-merger remnants by various GW observatories. 
As a rough approximation, we assume that a GW is detectable when its signal-to-noise ratio (SNR) is above $8$. 
We note, however, that the typical continuous-wave searches utilize different search methodology to the matched filtering (for example HMM), which may change the detectability slightly. 
However, because it may be challenging to use HMM to track vector modes that can evolve more rapidly than the scalar modes~\citep{Isi2019}, we consider standard matched filtering here (although we note that it is in principle possible to tune the short fourier transform length and coherent time in such a way as to accommodate the vector search). 
Therefore, the results her can be taken as an optimistic upper limit.
A search with real data will likely benefit from development of continuous-wave methodologies for the more rapidly evolving vector modes.

For the Fourier transformed amplitude $\tilde{h}(f)$, no analytical transformation exists and it is computationally expensive to perform numerical transforms. As a result, we apply the stationary phase approximation here (see Appendix \ref{app:A} for derivation):
\begin{equation}
    \label{eq:hf}
    \tilde{h}(f) \approx \frac{h_{0, peak}}{1+\frac{f-f_0}{2\dot{f}\tau_{GW}}} \sqrt{\frac{i}{2\dot{f}}} \textrm{exp}(-i2\pi\frac{(f-f_0)^2}{4\dot{f}}).
\end{equation}
Fig.\ref{fig:horizon} shows the horizon at different masses by LIGO at design sensitivity \cite{LIGOdesigned}, as well as next-generation ground-based GW observatories Cosmic Explorer \cite{Evans:2021gyd, cosmicexplorer,CE_2017}. 

We assume 3 years of observation time for scalar bosons and 1 day of observation time for vector bosons. We take $\chi=0.7$, a value similar to the spin of most observed merger remnants \cite{GWTC1, GWTC2, GWTC3}. We choose several values of $\alpha$ that would satisfy the "superradiance condition" specified in Eq. \ref{superradiance_condition}.

In Fig.\ref{fig:horizon}, we also show the scatter of merger population forecast obtained from \cite{Wierda2021}. 
As a summary of Ref.~\citet{Wierda2021}, the model uses the \textsc{Powerlaw+Peak} mass model fitted to the O1 and O2 observations with a merger-rate density tracing the star-formation rate density as predicted by Pop-I/II stars and population synthesis codes (we refer the interested reader to the article itself for details). 
The observed rate of mergers ($\textrm{SNR}_{\textrm{merger}}>8$) is $\sim 1900\ \textrm{yr}^{-1}$.  
Out of the $\sim 1900$ mergers with $\textrm{SNR}_{\textrm{merger}}>8$, four candidates clearly stand out in both scalar and vector cases. They are expected to host ULB clouds within the horizon of the LIGO detectors at design sensitivity ($\textrm{SNR}_{\textrm{cloud}}>8$), meaning it may be possible to observe signals from such clouds soon. 
For scalar boson, GW150914 cannot host ULB clouds within the detection horizon at design sensitivity. However, we may turn our eyes towards the future to see the possibility of having such detection. 
Note, however, that these estimates are based on theoretical modelling of the merger-rate density, where there is some variation between different population synthesis models~\citep{santoliquido2021cosmic}. 
A more accurate estimate may become feasible in the future, as the number of gravitational-wave detections grow. 

However for vector boson, we might be able to detect cloud hosted by the remnant of the observed merger: the first gravitation-wave observation event GW150914 was quite near and the remnant had very conducive environment for ultralight boson cloud detection \cite{ligo2016}. 
Indeed, it would be interesting to see if we would be able to detect the vector boson cloud hosted by the remnant of GW150914 in the observation run O1 data. 
Hence we also obtain the vector horizon at O1 sensitivity in Fig. \ref{fig:horizon_o1}. If ULBs exist in the right mass range, with $\alpha \sim 0.15-0.2$, which corresponds to a boson mass $\mu$ of $3.2 \times 10^{-13} - 4.3 \times 10^{-13}$, our results suggest that at O1 sensitivity may detect the boson cloud hosted by GW150914, or to partially rule out the relevant parameter space.

\section{\label{sec:analysis} Analysis}
\begin{table}
\centering
\caption{Properties of one of the interesting remnants expected to host boson clouds within the detection horizon at LIGO at design sensitivity: mass of the binary BH before coalescence $m_1$ and $m_2$, mass of the remnant $M$, spin of the binary BH before coalescence $\chi_1$ and $\chi_2$, spin of the remnant $\chi$, luminosity distance $d_L$, and SNR of the merger.}
\begin{tabular}{|c|c|c|c|c|c|c|c|} 
 \hline
 $m_1$ [$M_\odot$] & $m_2$ [$M_\odot$] & $M$ [$M_\odot$] & $\chi_1$ & $\chi_2$ & $\chi$ & $d_L [Mpc]$ & SNR\\ 
 \hline 
 28.67 & 20.51 & 49.19 & 0.45 & 0.25 & 0.7 & 160.17 & 19.53 \\
 \hline
\end{tabular}
\label{table:remnants_properties}
\end{table}

To illustrate how one can combine the merger information with the binary information, we choose a binary merger event that formed detectable bosonic clouds around merger remnants from our catalogue (Fig.~\ref{fig:horizon}). 
In particular, the parameters of the merger remnants are given in Table~\ref{table:remnants_properties}. 
For these two remnants, we simulate the GW from both the merger signal (using the \textsc{IMRPhenomPv2} waveform~\citep{lalsuite}) and the cloud signal (Eq.~\ref{eq:hf}). 
We then independently infer the binary and the cloud parameters using the \texttt{Bilby} nested sampling tool and combine the measurements together. 

We find that the merger signal and the cloud signal produce a single, consistent measurement of the boson cloud in the vector and scalar case (Fig.~\ref{fig:para_est}). 
Although we presume that the cloud waveform is precisely modelled, which is optimistic, the results demonstrate that the information from the boson cloud can be used to cross-verify the ultralight boson hypothesis in a way that is unique to merger remnants. 
For the scalar boson, the damping time $\tau_{GW}$ measurement cannot be well-measured as it is in the timescale of years, thus the peak is a bit off compared with other measurements. In contrast, the $\tau_{GW}$ for vector boson is in the timescale of days and can be well-measured, thus showing a more consistent peak with other measurements

We also perform the same analysis to see if we can disentangle the signal of scalar boson cloud from vector boson cloud. To do this, we inject the GW signal by scalar boson cloud and inference the boson mass using the vector model. In Fig. \ref{fig:disentangle}, we show that if a scalar boson signal injection is analyzed with the vector model hypothesis, we cannot obtain a consistent boson mass. Thus, our results suggest that we would be able to tell whether the signal comes from a scalar boson cloud or a vector boson cloud. 

To see how the strategy helps us to distinguish signals from ULB cloud and other CW sources, we inject a CW signal from pulsar and infer the signal using the boson cloud waveform. 
This ought to tell us whether or not it would be possible for other sources to mimick an ULB cloud signal. 
In Fig. \ref{fig:pulsar}, we show that if the signal is coming from a pulsar source, the inference would give inconsistent measurements of boson mass. In particular, we inject the signal from the Crab pulsar, with $f=59.25 \textrm{Hz}$ and $h_0=1.4\times10^{-24}$ , assuming it radiates at the spin-down limit, and a very small value of $\dot{f}$ and a very large value of $\tau_{GW}$ \citep{Creighton:2011zz} as an illustrative example.
Therefore by this analysis, we argue that it would be possible to distinguish ULB signals from other CW signals.\footnote{Though we also note that there may be other ways to discriminate between the different sources, for example GWs from spinning neutron stars typically decrease in frequency while GWs from boson clouds typically increase in frequency.}

To quantify the consistency of the measurements, we calculate the overlap Bayes factor following \cite{Haris:2018vmn}
\begin{equation}
    \label{eq:Bayes_factor}
    \mathcal{B} = \int \frac{P(\mu_1|d)P(\mu_2|d)P(\mu_3|d)P(\mu_4|d)}{P(\mu)^3} \,d\mu\,,
\end{equation}
which measures how much the measurements overlap. If the measurements are consistent, they would have a larger overlapping region, which corresponds to a larger Bayes factor. 
The prior probability $P(\mu)$ is set to be uniform at $10^{-15} - 10^{-10}$ eV, which corresponds to the frequency detectable by ground-based detectors (Eq. \ref{eq:f}), we set $\mu_i=\mu$, by definition, and $d$ stands for both data from the binary black hole merger and also the subsequent cloud waveform.
Although the Bayes factor can be quite sensitive to the prior information, our results show that one can robustly quantify the overlap and lack thereof. 
Indeed, in Table \ref{table:Bayes_factors}, we summarize the Bayes factor for each of the measurements. 
The Bayes factors for the scalar injection scalar hypothesis and vector injection vector hypothesis are large positive numbers due to the heavy overlap in posteriors. 
On the other hand, for the case of scalar signal injection with vector hypothesis, and  pulsar signal injection, the Bayes factor is zero due to the inconsistent measurements of boson mass. 

\begin{table}
\centering
\caption{Overlap Bayes factor of each signal analysed. We show the injected signal (left column), the hypothesis which we consider in estimating the Bayes factor (middle column), and the overlap Bayes factor (right column). The Bayes factor (Eq. \ref{eq:Bayes_factor}) quantifies how much each measurements overlap with each other. The more consistent the measurements are, the higher the Bayes factor. 
The Bayes factor correctly identifies the scalar/vector cloud when one is injected in the data and can also rule out the incorrect hypothesis when attempting to infer the parameters of a simulated waveform using a different hypothesis. 
}
\begin{tabular}{|c|c|c|}
\hline
Injection & Hypothesis & Overlap Bayes factor\\
 \hline
 Scalar & Scalar & $7.3\times10^9$\\
 \hline
 Vector & Vector  & $2.5\times10^{11}$\\
  \hline
 Scalar & Vector & 0\\
  \hline
 Pulsar & Scalar & 0\\
 \hline
\end{tabular}
\label{table:Bayes_factors}
\end{table}

\section{\label{sec:conclusions} Conclusions and Outlook}

We have quantified the horizon for finding boson clouds around merger remnants. By comparing with the existing forecast on merger events, we find that we may detect signals from boson clouds formed around merger remnants in the near future. In estimating the horizon distance, since no analytical transform is available, we use stationary phase approximation to perform the Fourier transform of the wave. To perform actual searches, dedicated continuous wave approaches and more sophisticated approaches would be needed. Furthermore, the estimates need to be interpreted with some care as they are to a degree subject to the changes in the merger-rate density, which suffer from uncertainty at high redshift~\citep[e.g.][]{santoliquido2021cosmic} (see also discussion in~\citep{Wierda2021}). Nevertheless, the current results outline a possible scenario. Interestingly, we also find that if ULBs are vector bosons and in the right mass range, the ground-based detectors may already have detected the signal of the boson cloud hosted by GW150914. 

Targeting merger remnants can indeed be a good strategy as it could provide a consistency test that might cross-verify the existence of ULBs. In particular, when the signal is produced by other CW sources, the measurement of boson mass will be inconsistent, which can help us to confirm if the source of the CW is the ULB cloud. Thus, the test would be able to disentangle between other continuous-wave sources and genuine ultralight boson signals at great accuracy. Furthermore, the same method would be able to disentangle between vector bosons and scalar bosons with great accuracy. This may provide a complementary strategy to other detection methods when we receive CW signals. Indeed, although targeting merger remnants may have the disadvantage of having a lower rate of detections compared to some of the alternatives~\citep{Isi2019}, it has the advantage of being able to robustly confirm that the signal indeed originates from an ultralight boson clouds. Since a robust verification of any new particle would likely require extraordinary evidence, a corroborating detection from a merger remnant in the scenario that ultralight bosons do exist would be quite valuable. 

In the future, the proof-of-concept analysis we have presented here would hopefully find applications to real data.  
To this end, future work may focus on building more sophisticated analyses targeting continuous waves with a rapid frequency drift, such as those expected from vector bosons. 
To this end, it will become more important to also include higher-order corrections to the gravitational-wave waveform and more agnostic search strategies; work towards a practical search strategy is being carried out by~\citet{WilliamInPrep}. 
Another important aspect in future analyses will also be to account for Earth's rotation, as the signals we target can last up to years. 
Nevertheless, the proof-of-concept analysis presented here demonstrates potential for an interesting consistency test with merger remnants.

\section*{\label{sec:acknowledgements} Acknowledgments}

We thank Richard Brito, Isaac C. F. Wong, William East and Ling Sun for useful comments and suggestions. 
O.A.H. is partially supported by grants from the Research Grants Council of the Hong Kong, The Croucher Foundation of Hong Kong and Research Committee of the Chinese University of Hong Kong. 
We acknowledge the software packages used, including \texttt{Matplotlib}~\cite{matplotlib}, \texttt{NumPy}~\cite{numpy}, \texttt{SciPy}~\cite{scipy}, \texttt{scikit-learn}~\cite{sklearn_api}, \texttt{Bilby}~\cite{bilby2019}, and \texttt{PyCBC}~\cite{pycbc}. The authors are grateful for computational resources provided by the LIGO Laboratory and supported by National Science Foundation Grants PHY-0757058 and PHY-0823459. This material is based upon work supported by NSF's LIGO Laboratory which is a major facility fully funded by the National Science Foundation.

\appendix
\section{\label{app:A} Stationary Phase Approximation}
In this appendix we provide the calculation of $\tilde{h}(f)$ using the stationary phase approximation.

The Fourier transform of the time-domain strain $h(t)$ is given by
\begin{align}
    \tilde{h}(f) &= \int h(t) e^{-i2\pi ft} dt  \label{eq:fourier} \\
    &= \int \frac{h_{0, peak}}{1+\frac{t}{\tau_{GW}}} \textrm{exp}(i2\pi(f_0+\dot{f}t+f)t) dt \\
    &= \int \frac{h_{0, peak}}{1+\frac{t}{\tau_{GW}}} \textrm{exp}(i2\pi\Phi(t)) dt
\end{align}

\noindent where phase $\Phi(t) = f_{0}t+\dot{f}t^2+ft$. The first derivative of phase $\Phi(t)$ is

\begin{equation}
\label{eq:phi_prime}
\Phi^{\prime}(t)= f_{0}+2\dot{f}t+f
\end{equation}

\noindent and the second derivative 

\begin{equation}
\Phi^{\prime\prime}(t)=2\dot{f}.
\end{equation}

For stationary phase approximation, the phase term is stationary when  $\Phi^{\prime}(t)|_{t=t_0}=0$. By \ref{eq:phi_prime}, it is equivalent to
\begin{equation}
t_0=\frac{f-f_0}{2\dot{f}}.
\end{equation}

Putting back $t_0$ into $\Phi(t)$ and $h_0(t)$
\begin{equation}
\Phi(t_0)=-\frac{(f-f_0)^2}{4\dot{f}}
\end{equation}

\noindent and

\begin{equation}
\label{eq:h0t0}
h_0(t_0)=\frac{h_{0, peak}}{1+\frac{f-f_0}{2\dot{f}\tau_{GW}}}.
\end{equation}

If we expand $\Phi(t)$ as a Taylor series about $t_0$ to the second order,

\begin{align}
\Phi(t)&\approx \Phi(t_0)+\frac{1}{2}\Phi^{\prime\prime}(t_0)(t-t_0)^2 \\
&= -\frac{(f-f_0)^2}{4\dot{f}}+\dot{f}(t-t_0)^2. \label{eq:phi_taylor}
\end{align}

Putting back the results of \ref{eq:phi_taylor} and \ref{eq:h0t0} to \ref{eq:fourier},

\begin{align}
    \tilde{h}(f) &= \int h_0(t_0) \textrm{exp}[i2\pi(\Phi(t_0)+\frac{\Phi^{\prime\prime}}{2}(t_0)(t-t_0)^2)] dt \\
    &= h_0(t_0) \textrm{exp}(-i2\pi\Phi(t_0)) \int \textrm{exp}(i2\pi\dot{f}(t-t_0)^2) dt \\
   &= \frac{h_{0, peak}}{1+\frac{f-f_0}{2\dot{f}\tau_{GW}}} \sqrt{\frac{i}{2\dot{f}}} \textrm{exp}(-i2\pi\frac{(f-f_0)^2}{4\dot{f}}) 
   \label{eq:h_final}
\end{align}

\noindent where in the final line we have used the result $\int \textrm{exp}(\frac{1}{2}icx^2)dx=\sqrt{2i\pi/c}$.

Using $\tilde{h}(f)$ in the form given in eq. \ref{eq:h_final}, we calculate the horizon when the SNR is $>8$. The SNR $\rho$ is given by

\begin{equation}
    \label{eq:snr}
    \rho = \sqrt{(h,h)} = \sqrt{4  \int_{0}^{\infty} \frac{|\tilde{h}(f)|^2}{S_{n}(f)}\ df }.
\end{equation}
\noindent where $S_n(f)$ is the power spectral density of the noise of a detector \cite{lalsuite}. 

\section{\label{app:B}Ultralight boson mass inference}
Since our consistency test is made using independent measurements of the ultralight boson masses, here we briefly recap the precise form of the independent boson mass measurements.
In particular, using the GW emission equations given in Sec. \ref{subsec: GW}, the boson masses 
\begin{equation}
\begin{split}
\mu_1^{(s)}&=\mu_1 (h^{(s)}_{0, peak}, M, \chi, d_L )\\
&= \frac{0.1c^3\hbar}{GM}\sqrt[7]{\frac{h^{(s)}_{0, peak}}{8 \times 10^{-28}}\frac{10M_{\odot}}{M}\frac{d_L}{Mpc}\frac{0.1}{\chi-\chi_f}}\\
\mu_2^{(s)}&=\mu_2 (f_{0}, M)\\
&=\frac{0.1c^3\hbar}{GM}\frac{f_0}{645Hz}\frac{10M_{\odot}}{M}\\
\mu_3^{(s)}&=\mu_3 (\dot{f}^{(s)}, M)\\
&= \frac{0.1c^3\hbar}{GM}\sqrt[19]{\frac{\dot{f}^{(s)}}{3\times10^{-14}Hz/s}(\frac{M}{10M_{\odot}})^2\frac{1}{\chi^2}}\\
\mu_4^{(s)}&=\mu_4 (\tau^{(s)}_{GW}, M, \chi)\\
&=\frac{0.1c^3\hbar}{GM}\sqrt[15]{\frac{6.5\times10^4yr}{\tau^{(s)}_{GW}}\frac{M}{10M_{\odot}}\frac{1}{\chi}}
\end{split}
\end{equation}
for the scalar boson case, and 
\begin{equation}
\begin{split}
\mu_1^{(v)}&=\mu_1 (h^{(v)}_{0, peak}, M, \chi, d_L )\\
&= \frac{0.1c^3\hbar}{GM}\sqrt[5]{\frac{h^{(v)}_{0, peak}}{4 \times 10^{-24}}\frac{10M_{\odot}}{M}\frac{d_L}{Mpc}\frac{0.1}{\chi-\chi_f}}\\
\mu_2^{(v)}&=\mu_2 (f_{0}, M)\\
&=\frac{0.1c^3\hbar}{GM}\frac{f_0}{645Hz}\frac{10M_{\odot}}{M}\\
\mu_3^{(v)}&=\mu_3 (\dot{f}^{(v)}, M)\\
&= \frac{0.1c^3\hbar}{GM}\sqrt[15]{\frac{\dot{f}^{(s)}}{1\times10^{-6}Hz/s}(\frac{M}{10M_{\odot}})^2\frac{1}{\chi^2}}\\
\mu_4^{(v)}&=\mu_4 (\tau^{(v)}_{GW}, M, \chi)\\
&=\frac{0.1c^3\hbar}{GM}\sqrt[11]{\frac{1 day}{\tau^{(v)}_{GW}}\frac{M}{10M_{\odot}}\frac{1}{\chi}}
\end{split}
\end{equation}
for vector boson. 
The four inferred boson mass $\mu$ can then used to cross-verify the existence of ultralight bosons.

\bibliographystyle{mnras}
\bibliography{bib.bib}

\end{document}